\documentclass[useAMS,usenatbib,usegraphicx]{mn2e}
\include{epsf}

\title[Chemodynamical analysis of bulge stars for simulated disc galaxies]
{Chemodynamical analysis of bulge stars for simulated disc galaxies}
\author[Rahimi et~al.]
 {A.~Rahimi$^{1}$\thanks{E-mail: ara2@mssl.ucl.ac.uk},
D.~Kawata$^{1}$
Chris B. Brook$^{2}$
Brad K. Gibson$^{2}$
\\
$^{1}$ Mullard Space Science Laboratory, University College London,
Holmbury St. Mary, Dorking, Surrey, RH5 6NT, UK
\\
$^{2}$Centre for Astrophysics, University of Central Lancashire, 
Preston, PR1~2HE, UK
}
\date{Accepted .
      Received ;
      in original form }

\pagerange{\pageref{firstpage}--\pageref{lastpage}}
\pubyear{2009}

\begin{document}

\maketitle

\label{firstpage}

\begin{abstract}
We analyse the kinematics and chemistry of the bulge stars of two simulated disc galaxies using our chemodynamical galaxy evolution code {\tt GCD+}. First we compare stars that are born inside the galaxy with those that are born outside the galaxy and are accreted into the centre of the galaxy. Stars that originate outside of the bulge are accreted into it early in its formation within 3 Gyrs so that these stars have high [$\alpha$/Fe] as well as having a high total energy reflecting their accretion to the centre of the galaxy. Therefore, higher total energy is a good indicator for finding accreted stars. The bulges of the simulated galaxies formed through multiple mergers separated by about a Gyr. Since [$\alpha$/Fe] is sensitive to the first few Gyrs of star formation history, stars that formed during mergers at different epochs show different [$\alpha$/Fe]. We show that the [Mg/Fe] against star formation time relation can be very useful to identify a multiple merger bulge formation scenario, provided there is sufficiently good age information available. Our simulations also show that stars formed during one of the merger events retain a  systematically prograde rotation at the final time. This demonstrates that the orbit of the ancient merger that helped to form the bulge could still remain in the kinematics of bulge stars. 
\end{abstract}


\section{Introduction}
\label{intro-sec}

In recent years, there has been progress in modelling disc galaxy formation in a Cold Dark Matter (CDM) Universe, using three dimensional numerical simulations \citep{nk92,sm94,bc00,anse03a,anse03b,bkgf04b,gwmb07,onb08}. Some studies include both Type II (SNe II) and Type Ia (SNe Ia) supernovae and discuss the details of the chemical properties in the simulated galaxies in an isolated halo collapse \citep{rvn96,b99} or hierarchical clustering \citep{bgmk05,rgm05}. These studies mainly focus on the disc component of the simulated galaxies. None, except \citet{nn03}, study the bulge formation in three dimensional chemodynamical simulations. \citet{nn03} suggest that the bulge may be composed of two chemically different components, one being formed through an early epoch subgalactic clump merger, and the other in the inner region of the disc after this merger. This paper focuses on the population of bulge stars built up through mergers, using higher resolution and more sophisticated chemodynamics simulations. Especially, we study the properties of accreted and locally formed stars within the bulge, which \citet{nn03} do not look at.

There are two major formation scenarios for bulge stars. The first scenario is secular evolution of the disc \citep[see, e.g.][]{cdfp90,nsh96,mn98}. In most secular evolution scenarios, gas and stars from the disc are funneled to the central regions of the galaxy. \citet{cdfp90} showed that it is possible to form a bulge via the secular formation of a bar, and \citet{mn98} showed that the disc of a young galaxy may become gravitationally unstable and fragment into large pieces. These large clumps fall into the central region due to dynamical friction and proceed to form a massive bulge via mergers. \citet{mn98} and \citet{iswg04} also suggest that several peculiar morphological structures seen in the Hubble Deep Field can be explained by a fragmented galactic disc model.  

\begin{table*}
 \centering
 \begin{minipage}{140mm}
  \caption{Simulation parameters}
  \begin{tabular}{@{}lllllllllll@{}}
  \hline
   Name & $M_{\rm vir}$ & $r_{\rm vir}$ & $m_{\rm gas}$ & $m_{\rm DM}$ & $e_{\rm gas}$ & $e_{\rm DM}$ & $\Omega_0$ & $h_0$ & $\Omega_{\rm b}$ \\
   & $(M_{\sun})$ & $(kpc)$ & $(M_{\sun})$ & {$(M_{\sun})$} & {$(kpc)$} & {$(kpc)$} & & & & \\
   \hline
Gal1 & $8.8\times10^{11}$ & 240 & $9.2\times10^{5}$ & $6.2\times10^{6}$ & 0.57 & 1.1 & 0.3 & 0.7 & 0.039  \\
Gal2 & $9.3\times10^{11}$ & 270 & $3.3\times10^{6}$ & $1.9\times10^{7}$ & 0.87 & 1.5 & 0.3 &
0.65 & 0.045 \\
 \hline
\end{tabular}
\end{minipage}
\end{table*}

The other major formation scenario for bulge stars is via mergers of subclumps, first shown by \citet{tt72}. Using numerical simulations and including star formation, gas dynamics and SNe feedback, \citet{nk92} showed that hierarchical clustering predicts a number of mergers at high redshift. In these merger processes, the number of bulge stars builds up by accreted stars which were already formed within the infalling galaxies, and by stars which formed locally within the main galaxy during and after the merger via starbursts. Although a number of studies focus on accreted stars \citep{bkgf03,hnnh06,fjbr06b}, there have been few studies comparing the properties of these accreted satellite stars with those of the stars formed locally in the galaxy. Recently \citet{zwbg09} compared the properties of accreted and locally formed stars in the halo of the galaxy. There are very few studies however, dedicated to the bulge chemodynamical properties.  This is the first study to compare accreted and locally formed stars within the bulge.

The chemical properties of stars are important to understand the formation history of the bulge. Chemical elements heavier than He are the end products of stellar evolution. The so-called $\alpha$-elements and iron (Fe) are of particular interest since it is known that they are produced primarily in SNe II and SNe Ia respectively. SNe Ia and SNe II have different timescales and thus studying the abundance ratios of the $\alpha$ elements with respect to Fe gives unique fossil information on the past conditions and evolution of the galaxy. 
Pure chemical evolution studies of the bulge such as \citet{mb90,fws03} are useful for reconstructing the star formation history of the bulge. \citet{mb90} studied the chemical abundances in the bulge, and looked at the [$\alpha$/Fe] ratios for different elements. They predicted a relative enhancement of [$\alpha$/Fe] in the bulge, due to the high Star Formation Rate (SFR). Based on the observed distribution of metallicities and using a simple star formation model, \citet{fws03} found that very short infall timescales are required in the bulge. These studies are useful to reconstruct the star formation history of the bulge. However, pure chemical evolution models do not answer how such a star formation history occurred in the galaxy, or predict any information about kinematics.

The findings of \citet{mb90} were confirmed by  the observational studies of \citet{mcr94}, \citet{fmcr06}, and \citet{zoclbh06}. By looking through Baade's Window, \citet{mcr94} found that the primary enrichment process of the bulge was SNe II. Recently, 8m class large telescopes have enabled observers to observe stars in the Milky Way bulge in more detail. Using a larger sample of stars at higher resolution \citet{zoclbh06} concluded that the Milky Way bulge likely formed through a short series of starbursts triggered by gas rich mergers in the early universe.
In the future, both kinematical and chemical properties of bulge stars will become available via projects such as APOGEE which can look through the obscuring dust to the bulge region of the Milky Way using near infrared multi-object spectrographs \citep{apms08}. ESA's Gaia mission should provide six dimensional phase space and chemical information on up to one billion stars in the Milky Way. These stars lie predominantly in the disc however there is some hope that Gaia may also observe significant numbers of bulge stars \citep{rrps05}. Ultimately, JASMINE, a planned Japanese space astrometry mission, will observe around ten million stars in the bulge by taking detailed astrometric readings in the infrared band \citep{gouda06}. This huge influx of new data is expected to spur new scientific results in different fields of astronomy. However, to extract the formation history of galaxies from such a large amount of observational data, theoretical models and predictions are crucial. Therefore, we initiate a project to study how we can tell the past formation history of the bulge through the current properties of bulge stars.

We study both chemical and kinematical properties of bulge stars using $\Lambda$CDM cosmological simulations. We take two high resolution simulations from the sample of \citet{bkg05}. In our $\Lambda$CDM simulation, bulges formed during hierarchical clustering of building blocks at high redshift. We therefore study the kinds of properties that are expected in bulge stars, if the bulge was formed through such a hierarchical clustering scenario. Therefore, our study does not mean to reproduce the Milky Way bulge which may or may not form through hierarchical clustering.

The outline for this paper is as follows. In Section 2 we describe our numerical simulations and define our bulge stars. In Section 3 we present the results of our chemodynamical analyses on accreted and locally formed stars  within the bulge. Finally we present our conclusions in Section 4.

\section{The code and model}
\label{code-sec}
To simulate our galaxies, we use our original galactic chemodynamics evolution code {\tt GCD+} developed by \citet{kg03a}. 
{\tt GCD+} is a three-dimensional tree $N$-body/smoothed particle hydrodynamics \citep[SPH;][]{ll77,gm77}  
code \citep{bh86,hk89,kwh96} that incorporates self-gravity, hydrodynamics, radiative cooling, star formation, supernova feedback, and metal enrichment. {\tt GCD+} takes into account chemical enrichment by both SNe II \citep{ww95} and SNe Ia \citep{ibn99,ktn00} and mass loss 
from intermediate-mass stars \citep{vdhg97}, and follows the chemical enrichment history of both the stellar and gas components of the system.

Radiative cooling, which depends on the metallicity of the gas
 \citep[derived with {\tt MAPPINGSIII}:][]{sd93} is taken into account. The cooling rate for a gas with solar metallicity is larger than that for gas of primordial composition by more than an order of magnitude. Thus, cooling by metals should not be ignored in numerical simulations of galaxy formation \citep{kh98,kpj00}. However, we ignore the effect of the UV background radiation for simplicity.

The two galaxies simulated here are from the sample of \citet{bkg05}, KGCD and AGCD, hereafter ``Gal1" and ``Gal2" respectively. Gal1 is a high resolution version of galaxy ``D1" in \citet{kgw04}. We used the multi-resolution technique in order to maximise the mass resolution within the regions where the disc progenitors form and evolve \citep{kgw04}. The initial conditions for Gal2 were kindly provided by M. Steinmetz, and are described in \citet{anse03a,anse03b}.

We summarise the properties of the galaxies in Table 1 taken from \citet{ckbtg06}. The first column represents the galaxy name; the second column, the virial mass; Column 3, the virial radius; Columns 4 and 5 represent the mass of each gas and DM particle in the highest resolution region, and Columns 6 and 7 are the softening lengths in that region.  The cosmological parameters for the simulation are presented in Columns 8--10. $\Omega_0$ is the total matter density fraction, $h_0$ is the Hubble constant (100 ${\rm kms^{-1}Mpc^{-1}}$) and $\Omega_{\rm b}$ is the baryon density fraction in the universe.
Note that the cosmology is slightly different between the two models, and the age of the Universe is 13.5 and 14.5 Gyrs for Gal 1 and Gal2 models respectively.

\begin{figure}
\centering
\includegraphics[angle=0,width=\hsize]{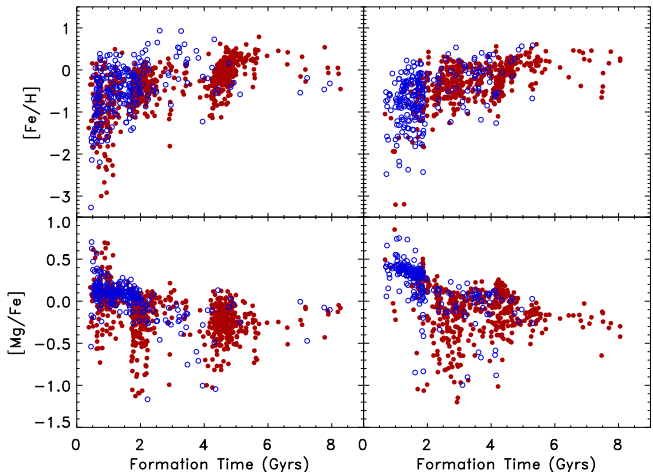}
\caption{
[Fe/H] (upper) and [Mg/Fe] (lower) against formation time ($t_{f}$) for Gal1 (left) and Gal2 (right). Red filled and blue open circles are in situ and accreted stars respectively. There are 3 major periods of star formation clearly distinguishable in Gal1. Star formation is smoother in Gal2.
}
\label{feoh-fig} 
\end{figure}

To identify the main progenitor galaxy, at each output of the simulation, a friend-of-friend (FOF) group finder is used to identify stellar groups. Specifying a linking length $b$ and identifying all pairs of particles with a separation equal to or less than $b$ times the mean particle separation as friends, stellar groups are defined as sets of particles connected by one or more friendship relations. The other main parameter in FOF is the minimum number of particles. By setting this parameter sufficiently high, one avoids including spurious objects that may arise due to chance. In our simulations we use a linking parameter $b=0.01$ and a threshold number of particles of 100. We define the largest group which has the highest number of the FOF identified particles as the main progenitor galaxy.

\begin{figure}
\centering
\includegraphics[width=\hsize]{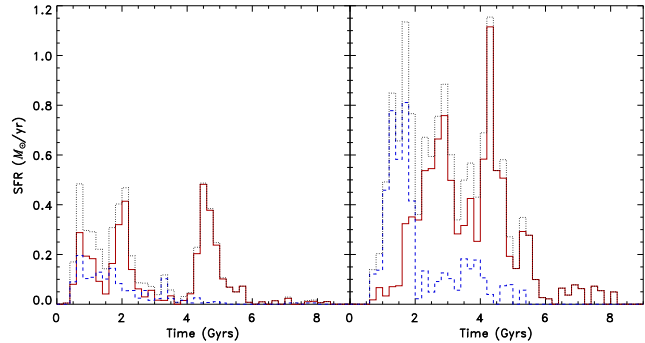}
\caption{
Global star formation rate (SFR) for Gal1 (left) and Gal2 (right). Red solid, blue dashed and black dotted are in situ, accreted and total stars respectively. 
}
\label{sfr-fig}
\end{figure}

Note that the version of {\tt GCD+} used in this paper \citep{kg03a} applies the SNe II yields calculated by \citet{ww95}. It is well known that this yield seems to be overestimated and leads to lower [$\alpha$/Fe], compared with that observed in low-metallicity stars in the solar neighbourhood \citep{tww95,bg97,glm97}. We adopted pure Woosley \& Weaver yields for Gal1. On the other hand, we apply a half Fe yield for Gal2. As a result, [Mg/Fe] values in Gal1 are underestimated compared to Gal2. Therefore, we cannot compare the absolute value of [Mg/Fe] between Gal1 and Gal2. However, we compare the relative difference of [Mg/Fe] among different samples of stars within a simulated galaxy.
Our code does not ``disperse" metals between gas particles. Therefore, the spread in the metallicity distributions will be artificially high, while the peak of the distributions (the average) should be robust (or at least the distances between the peaks). 

We identify bulge particles using the simulation output at the latest time. For Gal1, we use the output at $z=0.1$, as going to any lower redshift results in an unacceptable amount of contamination from low-resolution particles in our simulated galaxy. For Gal2, we use the output at $z=0$.
To define the bulge, we first set the disc plane of our galaxies to be the x-y plane and the rotation axis to be the z-axis. The bulge was defined as the central 2 kpc of the galaxy, excluding $\pm1$ kpc in the z-direction due to contamination with disc stars. Avoiding the disc plane will provide us with a ``cleaner" sample since the central region of the disc suffers from ``overcooling" problems and therefore produces too many stars throughout the simulations. All stars found within the remaining volume described above we refer to as  ``bulge stars'', and we analyse their properties in the next section.

\section{Results}
\label{res-sec}

Fig.~\ref{feoh-fig} shows [Fe/H] (upper panel) and [Mg/Fe] (lower panel) as a function of the time ({$t_f$}) when bulge stars in our two galaxies formed. There is clearly a different formation history for the bulges of the two galaxies. The bulge of Gal1 forms via a series of three mergers within the first 5 Gyrs. The bulge of Gal2 is formed in a similar timescale, however without the distinctly separated star formation epochs. Star formation here is more smooth and continuous. 
Both galaxies formed through hierarchical clustering, and bulge formation is associated with a series of mergers at early epochs. However, Gal1 and Gal2 have different merger histories and therefore star formation histories. The aim of this study is to find out how we can tell such a difference in merger histories at a high redshift from the current properties of bulge stars. To this end, first we separate the bulge stars into accreted stars and locally formed stars, and compare their properties in Section~\ref{comp-acc-ins}. From the current chemical and kinematical properties of bulge stars, we study how we can distinguish accreted stars from locally formed stars. In Section~\ref{age-analysis} we compare the properties of the bulge stars with different formation epochs to see whether the memory of mergers at different epochs persists or not.



\subsection{Accreted and in situ stars}
\label{comp-acc-ins}

In this section, we compare the chemical and kinematical properties of accreted and locally formed stars. We trace back the formation time and location for all the bulge stars, and any stars that are formed locally within a radius of 5 kpc from the largest progenitor are given the title ``in situ stars". 
Stars that form at a radius greater than 5 kpc from the centre of the galaxy and end up within the area defined above as the bulge, we term ``accreted stars". We chose 5 kpc arbitrarily. We experimented using a larger radius up to 20 kpc, and generally found the same conclusion.

\begin{figure}
\centering
\includegraphics[width=\hsize]{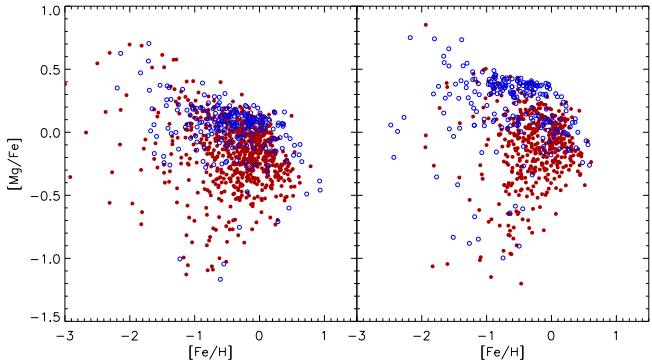}
\caption{
[Mg/Fe] vs [Fe/H] for Gal1(left) and Gal2 (right). Red filled and blue open circles are in situ and accreted stars respectively. As expected from Fig.~\ref{feoh-fig} accreted stars have generally higher [Mg/Fe] ratios. A population of accreted stars in Gal2 have high [Mg/Fe] because these stars formed in a large satellite which merged into the main galaxy at an early epoch.
}
\label{mgfevsfeh-fig}
\end{figure}

Fig.~\ref{feoh-fig} shows [Fe/H] and [Mg/Fe]  vs formation time for accreted and in situ stars with different symbols. 
Mg is one of the so-called $\alpha$-elements. These elements are primarily produced when massive stars with short lifetimes explode as SNe II. Fe is produced predominantly in SNe Ia, from lower mass binary stars with longer lifetimes. Accreted stars have lower [Fe/H] and higher [$\alpha$/Fe] since they form in early epochs before the enrichment from SNe Ia becomes important. In situ stars continue to undergo strong star formation for up to 5 Gyrs (as seen in Fig.~\ref{sfr-fig}, which shows the SFR history for in situ and accreted stars).
Fig.~\ref{mgfevsfeh-fig} shows [Mg/Fe] vs [Fe/H]. Note that there exists a distinct population of accreted stars in Gal2 with high [$\alpha$/Fe]. This is because these accreted stars were formed in a large companion galaxy at early epochs ($t$ $<$ 1.5 Gyrs) that underwent a major merger with the main galaxy (Fig.~\ref{sfr-fig}). This demonstrates that the probability distribution of high [Mg/Fe] stars may be sensitive to merger history at high redshift.

\begin{figure}
\centering
\includegraphics[width=\hsize]{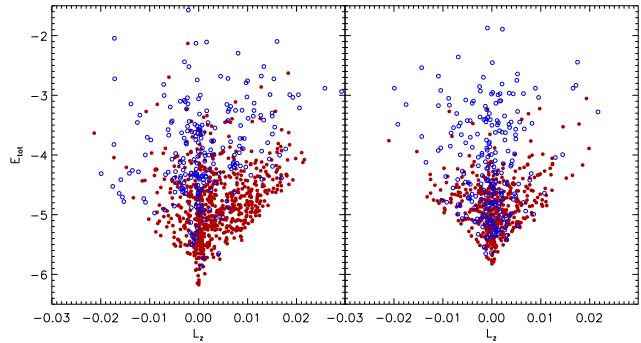}
\caption{
Total energy ($E_{\rm tot}$) vs angular momentum ($L_z$) for Gal1(left) and Gal2 (right) for accreted (open blue circles) and in situ (filled red circles) stars. Accreted stars have higher $E_{\rm tot}$. For Gal2 the distinction is unclear. This is due to a major merger early in its history bringing stars of a satellite galaxy into the main galaxy as accreted stars.
}
\label{etotvslz-fig}
\end{figure}

We find that Fig.~\ref{etotvslz-fig} is the most useful for identifying the accreted stars. It shows the stars total energy ($E_{\rm tot}$) plotted against their $z$-component of angular momentum ($L_z$) which is also used to identify accreted stellar groups in halo stars \citep{hdz00,kgkg05,fjbr06b}. The accreted stars have a relatively higher $E_{\rm tot}$ and are spread to a large range of $L_z$. This is especially the case for Gal1. For Gal2 however, the distinction is less clear. There exists a significant population of accreted stars with low $E_{\rm tot}$ in Gal2. This is due to the major merger this galaxy experienced at an early epoch in its history. These accreted stars with low $E_{\rm tot}$ were actually formed in the centre of a large satellite galaxy that merged with the main galaxy early on which explains why their $E_{\rm tot}$ is comparable to the in situ stars. 

We conclude that the $E_{\rm tot}$ - $L_z$ diagram is a powerful tool to identify accreted stars in the bulge. Its use is known for halo stars (Brook et al 2003, Morrison et al 2009). Our study shows that it can be applied for bulge stars which experienced more violent merger histories. However there are significant overlaps with in situ stars. Especially, it is difficult to identify stars which are born in a large satellite galaxy that merged at an early epoch.

\subsection{Multiple merger bulge formation}
\label{age-analysis}

The left panel of Fig.~\ref{sfr-fig} clearly shows that Gal1 has 3 bursts of star formation. From snapshots, we confirmed that each of the bursts corresponds to a merger. We therefore divided the bulge stars into 3 samples corresponding to each of the bursts. The  stars formed during the period of the first burst ($t_{f}$ $<$ 1.5 Gyrs) are referred to as ``old" bulge stars; those formed during the second burst (1.5 $<$ $t_{f}$ $<$ 3 Gyrs) are called ``intermediate" bulge stars; and those formed during and after the third and largest burst  ($t_{f}$ $>$ 3 Gyrs) are called ``young" bulge stars. From the right panel of Fig.~\ref{sfr-fig} we see that Gal2 also has three bursts of star formation, although there is significant star formation between the bursts. Therefore, we also assign 3 periods of star formation for Gal2. Our definition for the 3 time periods here was slightly different to Gal1, reflecting the different star formation history in Gal2. The formation periods of old, intermediate and young stars for Gal2 are defined as $t_{f}$ $<2$ Gyrs, 2 $<$ $t_{f}$ $<$ 4 Gyrs, and $t_{f}$ $>$ 4 Gyrs respectively.  

\begin{figure}
\centering
\includegraphics[angle=0,width=\hsize]{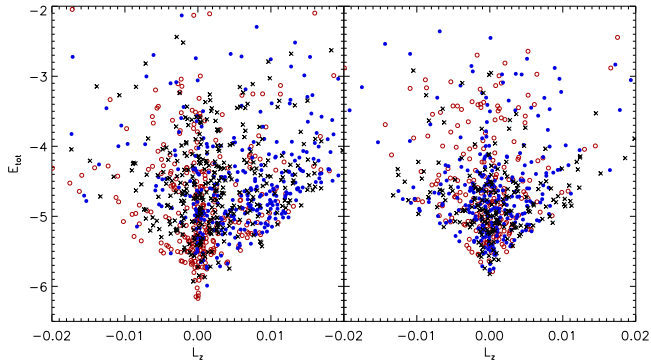}
\caption{
As in Fig.~\ref{etotvslz-fig}, but now using different aged bulge stars for Gal1 (left) and Gal2 (right). Red open circles, blue filled circles and black crosses are old, intermediate and young bulge stars respectively.  Trends are visible, especially in Gal1 which contains an intermediate aged population with a distinctly high $L_z$, possibly indicating the prograde orbit of a merger at that epoch.
}
\label{ageetotvslz-fig}
\end{figure}

In Fig.~\ref{ageetotvslz-fig} we replot $E_{\rm tot}$ vs $L_z$ for the three different groups of bulge stars. In Gal1 (left panel), the old stars are generally clustered about the centre with low $\vert$$L_z$$\vert$ and $E_{\rm tot}$. Young stars have higher $E_{\rm tot}$ and are less clustered about low values of $\vert$$L_z$$\vert$. In Gal1 we clearly see that the majority of intermediate stars have high $L_z$. This indicated that the intermediate population may have resulted from a specific merger and are orbiting in a prograde direction.
This demonstrates that some information from the ancient history of mergers can persist in $E_{\rm tot}$ and $L_z$ properties. If then, in addition, the ages of the stars are measured very accurately, the diagram of $E_{\rm tot}$ and $L_z$ may become a powerful tool to reconstruct the merger history. The picture for Gal2, again is less clear. The three populations in Gal2 are well spread on the diagram. No group occupies a distinct region.

\begin{figure}
\centering
\includegraphics[width=\hsize]{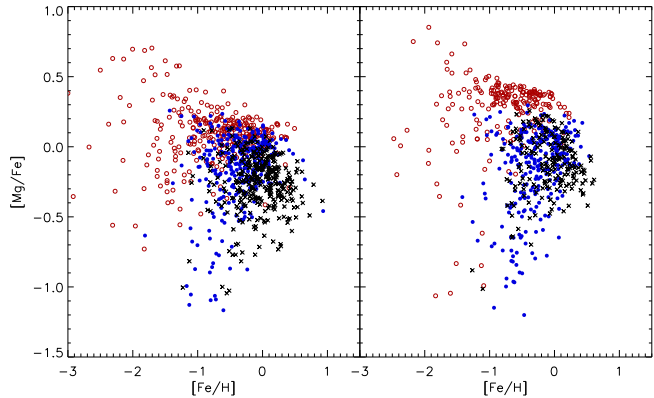}
\caption{
As in Fig.~\ref{mgfevsfeh-fig}, but now using different aged bulge stars. Red open circles, blue filled circles and black crosses are old, intermediate and young bulge stars respectively. For Gal1 again, the distinction between the different groups is clearer.
}
\label{agemgfevsfeh-fig}
\end{figure}

\begin{figure}
\centering
\includegraphics[width=\hsize]{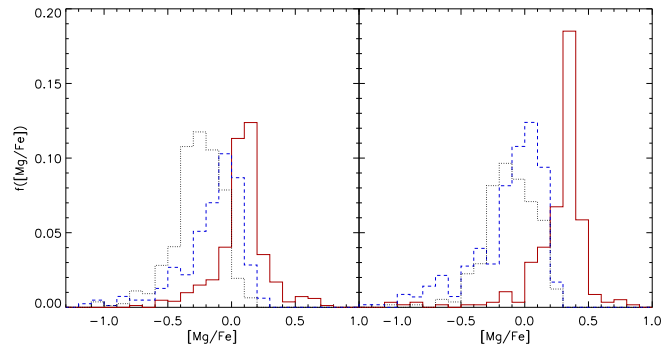}
\caption{
Histogram of [Mg/Fe] for different age samples of bulge stars for Gal1 (left) and Gal2 (right). Red solid, blue dashed and black dotted are old, intermediate and young bulge stars respectively. 
}
\label{hist-fig}
\end{figure}

Fig.~\ref{agemgfevsfeh-fig} shows that the old stars have the highest [Mg/Fe]. The younger stars have lower [Mg/Fe] and a higher metallicity as expected. This is because at later times the contribution from SNe Ia has become significant. The pattern is less clear for Gal2, reflecting its more continuous formation history, which diffuses out any trends. We plot the 3 samples of bulge stars as a probability distribution function for [Mg/Fe] as in Fig.\ref{hist-fig}. We see that they each occupy a separate distinct region, although there is some overlap. 
As mentioned in Section 2, our simulations do not take into account metal mixing between the gas particles. We overestimate the scatter of metal abundance and abundance ratios for star particles. Therefore, if the correct metal mixing is taken into account, [Mg/Fe] distributions of these 3 groups of stars will be more clearly different.
Note however that here we have assumed that we have accurate age information available for bulge stars. However, in reality, with the current facilities, it is difficult to obtain the ages of the bulge stars. Our study demonstrates that if the accurate age information becomes available (as it will in the future), we can identify bulges formed through distinct multiple mergers using the age-[Mg/Fe] relation (Fig.~\ref{feoh-fig}). Then, for bulge stars that are formed as a result of mergers at different epochs, we would expect differing values for [Mg/Fe], and for younger stars to have lower [Mg/Fe]. In addition, the $E_{\rm tot}$ - $L_z$ diagram can give us some information on the orbits of merging building blocks.

\section{Summary} 
\label{conc}

In this study we have analysed the chemistry and the dynamics of the bulge stars of two simulated disc galaxies. The two galaxies had different formation histories and were simulated under different cosmologies. However both galaxies were similar in mass and size to the Milky Way, and contained distinct gas and stellar disc components \citep{bkg05,ckbtg06}.

We started by defining the bulge region to be the central 2 kpc of the galaxy excluding $\pm1$ kpc in the z-direction to avoid contamination with disc stars. In this study we compared the properties of accreted and in situ stars within the bulge. Since accreted stars tend to form in early epochs, they have lower [Fe/H] and higher [Mg/Fe] ratios. We also found that even within the bulge, the accreted stars tend to have higher $E_{\rm tot}$ although there is significant overlap with in situ stars. We found that accreted stars formed in a large satellite galaxy that merged into the main galaxy can have low $E_{\rm tot}$ and are difficult to distinguish from in situ stars.

The bulge of one of the simulated galaxies was formed via a series of three distinct mergers in the first 5 Gyrs as confirmed by snapshots of the simulations. We found that such multiple star formation bursts will show up in the age-metallicity relation or more evidently in the age-[Mg/Fe] relation, provided there is accurate age information available. Then, stars that formed in earlier mergers tend to have higher [Mg/Fe], and different mergers have distinctively different [Mg/Fe]. Therefore, the probability distribution of [Mg/Fe] is sensitive to the number of mergers and the epochs of the mergers. We found that even in the bulge $E_{\rm tot}$ and $L_z$ information persists reasonably well till $z=0$. Therefore, if we could pick up stars formed in a merger at a specific epoch using the age information, $E_{\rm tot}$ and $L_z$ would be able to tell us about the orbital information of the merger, like whether it was in a prograde or retrograde direction.

The other simulated galaxy had undergone a major merger at a very early epoch, followed by a series of minor mergers between two later major mergers, which led to a smoother star formation history. This meant that much of the chemodynamical information had been smeared out. We found that a significant number of accreted bulge stars had high values of [Mg/Fe], due to significant star formation in two large progenitors at early epochs. This sensitivity to the star formation rate at early epochs means that finding such bulge stars with high values for [Mg/Fe] could depend on the size of the progenitors at the very early epochs.

Despite differences, our two simulated galaxies share some common formation features. There is evidence for the bulge of both galaxies forming via three major bursts of star formation. This is because in hierarchical clustering scenarios, bulges are likely to form through multiple rather than single mergers. If then, such merger epochs are separated by enough time, we may obtain information on these mergers from the properties observed in our simulated galaxies at $z=0$.

It is not yet possible from the current observations to obtain accurate age information for bulge stars. However in the near future, when increasingly accurate age information becomes available from combinations of the near infrared astrometry satellite JASMINE, and high resolution spectroscopy in the optical \citep{fmcr06} and near infrared with APOGEE \citep[see][]{apms08}, our findings will prove useful for identifying a possible scenario for the past formation history of the Galactic bulge.


\section*{Acknowledgments}

AR acknowledges the support of the UK's Science \& Technology
Facilities Council (STFC Grant) and would also like to thank Curtis Saxton and Samantha Oates for their help with IDL, and Curtis Saxton for his help in proof-reading the paper.

\bibliographystyle{mn}
\bibliography{dkref}

\label{lastpage}

\end{document}